\documentstyle[12pt]{article}
\def\Journal#1#2#3#4{{#1} {\bf #2}, #3 (#4)}

\def\NIMA{{\em Nucl.\ Instrum.\ Methods}\ A}

\def\NPA{{\em Nucl.\ Phys.}\ A}
\def\PLB{{\em Phys.\ Lett.}\ B}
\def\PRL{\em Phys.\ Rev.\ Lett.}
\def\PRD{{\em Phys.\ Rev.}\ D}
\def\PRC{{\em Phys.\ Rev.}\ C}
\def\ZPA{{\em Z.\ Phys.}\ A}
\begin{document}
\begin{center}
{\Large\bf Saturnales\ \footnotetext{Concluding talk at ``les 20 ans de
Saturne-2'' colloquium, Paris, 5 May 1998}}\\[3ex]
{\large Colin Wilkin}\\[1ex]
{\normalsize University College London, London, WC1E 6BT, UK}\\[5ex]
\end{center}

\centerline{\bf R\'esum\'e}
\begin{verse}
{\it Un choix personnel a \'et\'e fait d'\'ev\'enements m\'emorables du 
programme de physique r\'ealis\'e \`a Saturne-2 pendant les 
derniers 20 ans.}\\[2ex]
\end{verse}

C'est la fin, la fin de la r\'eunion et la fin de Saturne, et les gentils
organisateurs m'ont demand\'e de dire quelques mots sur les deux
\'ev\'enements.
Nous nous sommes r\'eunis ici \`a Paris pendant ces deux derni\`eres journ\'ees
afin de c\'el\'ebrer le travail de tout un ensemble de gens et d'appareils
group\'es autour d'un acc\'el\'erateur, Saturne-2, qui a \'et\'e arr\^et\'e en
d\'ecembre.

Je ne vais pas prononcer un discours politique mais il y a une mise au point
importante \`a faire. La facilit\'e NSF \`a Daresbury fut ferm\'ee
dans les ann\'ees 80, \`a la mesure de sa productivit\'e, par un
comit\'e de physiciens confront\'e
\`a une crise budg\'etaire caus\'ee par le besoin de soutenir la physique des
particules \'el\'ementaires, \`a l'\'epoque o\`u le franc suisse se trouvait en
hausse. Quand les physiciens nucl\'eaires mirent en question cette d\'ecision,
 l'administration leur r\'epondit qu'elle fut prise pour des raisons
de physique, ce qui \underline{n'est pas} la m\^eme chose qu'une d\'ecision
prise par des physiciens
sous la contrainte des circonstances! En cons\'equence de cette confusion
d\'elib\'er\'ee, qui fut transmise \`a la presse, les physiciens
nucl\'eaires perdaient non seulement leur machine mais ils se sentirent
insult\'es et d\'evalu\'es par dessus le march\'e. Ne
vous laissez donc conter par personne que
Saturne n'\'etait plus capable de continuer faute de probl\`emes
int\'eressants \`a \'etudier, ou par manque de comp\'etitivit\'e au niveau
international pour y arriver. Les deux seraient faux!

La flexibilit\'e de l'acc\'el\'erateur, avec son choix large de types de
particules et d'\'energie et une souplesse de r\'eglage en \'energie,
combin\'ee avec une collection de spectrom\`etres et polarim\`etres sans
pareille, en faisait un outil splendide pour d\'enouer des probl\`emes dans les
diff\'erents domaines de la physique. N\'eanmoins cette installation fut
ferm\'ee. Disons seulement que les autorit\'es avaient leurs priorit\'es
ailleurs que les n\^otres.

Chacun des excellents orateurs a \'et\'e oblig\'e de faire une s\'election,
 dans le domaine de son expos\'e, en partant d'un mat\'eriau d'une grande
richesse, et en une petite demi-heure, j'essaierai de choisir, dans l'ensemble
du programme du laboratoire, quelques \'ev\'enements m\'emorables pour moi
personnellement, tout en essayant de montrer comment, dans de nombreuses 
circonstances, ils ont influenc\'e les d\'eveloppements dans d'autres
instituts de recherche. Quoique Saturne soit mort, son g\'enome survit.

Avant de commencer, laissez-moi encore ajouter une remarque politique. Il est
bien plus facile d'obtenir de l'argent pour des exp\'eriences nouvelles
plut\^ot que de terminer les analyses des donn\'ees prises \`a Saturne.
 Aussi est-il plus agr\'eable
pour les physiciens de se lancer avec enthousiasme dans des
propositions nouvelles. N\'eanmoins, nous connaissons tous beaucoup
d'exp\'eriences dont les donn\'ees se trouvent encore sur bande et qui
repr\'esentent des ph\'enom\`enes int\'eressants mais qui risquent de prendre
la poussi\`ere dans les armoires sans \^etre exploit\'ees. Je ne veux
embarrasser personne en citant des exemples, au moins en public, mais je fais
appel aux
physiciens et \`a leurs directeurs de laboratoires pour mettre de c\^ot\'e un
peu de temps ainsi que de l'argent pour extraire le maximum des donn\'ees
prises \`a Saturne.

Il y a exactement 30 ans, j'ai fait mon premier discours \`a Saclay, et
en mai 68 les gens se pr\'eoccupaient de bien d'autres choses que de
l'application des mod\`eles du type Glauber \`a la diffusion des protons de 1 
GeV par les noyaux de carbone ou bien d'oxyg\`ene, qui faisait le sujet de
mon discours. Malgr\'e le fait que l'acc\'el\'erateur Saturne a \'et\'e
enti\`erement reconstruit il y a 20 ans, la continuit\'e du
programme de physique, ainsi que de certains \'equipements permet de 
m'\'egarer au-del\`a de la fronti\`ere de 1977. Comme je ne suis pas un
``homme de machine'' je cours donc le risque de ne pas insister suffisamment sur
le r\^ole de l'\'equipement dans ma  p\'eroraison. Pour \'eviter ce danger je
vous recommande l'excellente "vulgarisation" sur le travail realis\'e \`a
Saturne~\cite{Saturne}, m\^eme si vous \^etes d\'e\c{c}us de ne pas y
trouver un compte-rendu des bonnes et mauvaises ann\'ees! Si vous avez re\c{c}u
la version anglaise elle ne compte que 18 pages tandis que l'original en
fran\c{c}ais en a 48.

On n'oubliera pas non plus les comptes-rendus des Journ\'ees d'Etudes Saturne,
 \'edit\'es pendant des ann\'ees par Pierre Radvanyi et
Mme~Bordry. Ces volumes
constituent une v\'eritable mine d'informations utiles, sur Saturne, son
\'equipement et sa physique et repr\'esentent ainsi une chronique fid\`ele des
r\'eunions qui jou\`erent un r\^ole si important dans le d\'eveloppement de la
culture unique de Saturne. A titre d'exemple, mes premi\`eres id\'ees sur
la possibilit\'e d'un \'etat quasi-li\'e $\eta\,^3$He me sont venues
pendant les discussions au Mont Sainte-Odile, et les premiers calculs
furent faits dans le train de Strasbourg \`a Paris. Les autres gens
se souviennent plut\^ot des banquets et de la diff\'erence entre Roscoff
et Cavalaire.

La diffusion \'elastique proton-noyau et l'excitation des nivaux nucl\'eaires
furent les premiers sujets o\`u SPESI se fit remarquer sur le plan
international. Comme discut\'e par Vorobyov, ce programme \'etait une extension
du travail ant\'erieur de Palevsky et ses amis au Cosmotron.
Le discours de Vorobyov m'a rappel\'e les calculs de ma jeunesse mais
j'avais compl\`etement oubli\'e que le programme de la collaboration
Gatchina-Saclay \'etait si vaste. N\'eanmoins c'\'etait un programme de
physique nucl\'eaire classique o\`u les seuls degr\'es de libert\'e
\'etaient ceux des nucl\'eons. Dans le mod\`ele de Glauber ils ont obtenu
des densit\'es de mati\`ere dans plusieurs noyaux, et aussi des
densit\'es de transition, sans y mettre une forme sp\'ecifique de la
densit\'e du noyau. Donc ils ont pu regarder les changements de densit\'e
syst\'ematiques d'un isobare \`a un autre.

Puisqu'on
a \'evoqu\'e l'influence de Palevsky sur la construction de Saturne-1,
permettez-moi d'ajouter aussi un peu d'histoire puisque je travaillais
avec lui \`a Brookhaven en 1966-67. Saturne-1 \'etait une copie
fid\`ele du Cosmotron, et Harry n'avait jamais compris pourquoi
les Fran\c{c}ais avaient tenu \`a prendre les plans originaux pour
faire leur machine et ne voulaient pas introduire les am\'eliorations
\'etudi\'ees entre-temps par les ``machinistes'' am\'ericains. Par ailleurs si
Palevsky \'etait connu pour la diffusion \'elastique, sa vraie
passion \'etait plut\^ot pour une comparaison avec les spectres inclusifs pour
essayer d'obtenir des effets de corr\'elation dans le noyau \`a partir
des r\`egles de somme.

Le premier
mat\'eriau vraiment innovant fut une investigation des r\'eactions de
pick-up aux \'energies interm\'ediaires. SPESI \'etait capable de mesurer avec
facilit\'e la distribution de $^{12}$C$(p,d)^{\,11}$C$^*$ pour un demi-douzaine
d'\'etats excit\'es de $^{11}$C \`a 700~MeV~\cite{Thirion}. C'est une
co\"{\i}ncidence heureuse si Jacques Thirion a pr\'esent\'e ces premiers
r\'esultats \`a la m\^eme conf\'erence o\`u H\"oistad montrait les donn\'ees
d'Uppsala sur la r\'eaction $(p,\pi^+)$, puisque ces deux processus physiques
sont \'etroitement li\'es. A la diff\'erence des diffusions \'elastiques
proton-noyau \`a petit angle, ces r\'eactions \`a grand transfert
d'impulsion ont un lien intime avec les degr\'es de libert\'e du pion ou
du $\Delta$ dans les noyaux. Cette importance des m\'esons virtuels ou
des isobares, m\^eme pour des r\'eactions o\`u il n'y a pas de m\'esons
ni dans l'\'etat initial ni dans l'\'etat final, est quelque chose dont nous
devions tenir compte dans beaucoup d'exp\'eriences \`a Saturne-2. A titre de
simple remarque, les \'ev\'enements rares en r\'eactions hadroniques ne sont
jamais aussi rares qu'on l'imagine sur la base des degr\'es de
libert\'e purement nucl\'eoniques dans le noyau.

Pour moi personnellement le moment le plus excitant au LNS a \'et\'e la
nuit o\`u
nous avons mesur\'e la production du m\'eson $\eta$ dans la r\'eaction
$\vec{d}p\to\, ^3$He$\,\eta$ pr\`es du seuil \`a SPESIV~\cite{Berger}, m\^eme
si elle avait \'et\'e \'etudi\'ee auparavant \`a des \'energies plus
\'elev\'ees~\cite{Berthet}. Ce fut \'egalement un moment important pour le
laboratoire car les exp\'eriences \`a SPESII sur les
d\'esint\'egrations rares~\cite{Mayer1} en r\'esult\`erent.

\begin{itemize}
\item
Les r\'esultats ont montr\'e une section efficace d'une grandeur inattendue,
et cela est probablement d\^u \`a un faisceau pionique virtuel,
cr\'e\'e sur un des nucl\'eons dans le deuton, le pion produisant un m\'eson
$\eta$ sur le deuxi\`eme nucl\'eon de la cible. De telles id\'ees sont
maintenant utilis\'ees dans l'interpr\'etation de la production de m\'esons
au seuil dans les collisions d'ions lourds.
\item
La variation brusque de la section efficace pr\`es du seuil implique
l'existence d'un \'etat quasi-li\'e $\eta\,^3$He, une forme nouvelle de
la mati\`ere nucl\'eaire \`a haute \'energie d'excitation,
 dont l'existence a \'et\'e nettement confirm\'ee pour le syst\`eme
$\eta\,^4$He~\cite{Frascaria,Willis}.
\item
Ce dernier fournit une explication naturelle pour la violation de la
sym\'etrie de charge en $dd\to\, ^4$He$\,\pi^0$, dont un signal avait \'et\'e
observ\'e par le groupe inoubliable ER54 un peu en-dessous du
seuil de production du m\'eson $\eta$~\cite{Goldzahl}.
\item
Cela avait permis la mesure la plus pr\'ecise au monde de la masse du
m\'eson $\eta$~\cite{Plouin1}.
\item
L'exp\'erience avait ouvert une gamme de possibilit\'es pour \'etudier, avec
des faisceaux \'etiquet\'es de $\eta$, les d\'esint\-\'egrations rares de ce
m\'eson. Des r\'esultats importants sur les d\'ecroissances
$\eta\to \gamma\gamma$ et $\eta\to \mu^+\mu^-$ ont \'et\'e obtenus
~\cite{Mayer2}. Il est profond\'ement regrettable qu'on ait d\^u interrompre
d\'efinitivement ce programme d'\'etude des d\'esint\'egrations rares
\`a SPESII pour des
questions pratiques et financi\`eres. Quelle catastrophe!
\item
Les r\'esultats ont sucit\'e les \'etudes des \'etats plus lourds dans le
continuum en $pd\to\,^3$He$\,X$ dans les conditions dites du
seuil~\cite{Plouin2}, et en particulier  l'\'etude des m\'esons $X$  tels
que les $\omega$~\cite{Ralf1}, $\eta'$ et $\phi$~\cite{Ralf2}.
\item
Ce fut aussi l'inspiration pour les gens qui ont \'etudi\'e la production
au seuil du $\eta$
\cite{Bergdolt,Pinot}, $\omega$~\cite{Hibou2}, $\eta'$~\cite{Hibou1} et
$\phi$~\cite{Disto} en diffusion nucl\'eon-nucl\'eon, utilisant diverses
 techniques exp\'erimentales.
\end{itemize}

Presque tous ces th\`emes, n\'es dans une seule nuit d'efforts
passionn\'es, seront maintenant exploit\'es dans d'autres sites d'exp\'eriences,
 alors que les irr\'eductibles pr\'etendront, peut-\^etre \`a juste titre, qu'on
aurait pu faire encore mieux \`a Saturne! Donc il y a un grand programme de
d\'esint\'egrations rares \`a Uppsala et des recherches pour des \'etats
quasi-li\'es $\eta$-noyau \`a J\"ulich (et pendant la r\'eunion Paul Kienle
m'a expliqu\'e qu'il y avait des propositions similaires aussi \`a GSI).
Les deux laboratoires vont faire des \'etudes de production du
syst\`eme $\eta^{\,3}$He et chercher une \'eventuelle brisure de
la sym\'etrie de charge dans la r\'eaction $dd\to ^{\,4}$He$\,\pi^0$. J'ai
m\^eme entendu dire qu'en 2003 Fran\c{c}ois Plouin aura fini son analyse
de la masse du $\eta$, mais je n'en ai pas encore re\c{c}u confirmation.

Saturne \'etait renomm\'e au niveau international pour l'intensit\'e et la
qualit\'e de ses faisceaux de protons et deutons polaris\'es~\cite{Saturne},
 gr\^ace \`a la combinaison de la source d'ions polaris\'es Hyperion, du
mini-synchroton Mimas servant comme injecteur, et de l'\'etude
minutieuse des r\'esonances d\'epolarisantes \`a Saturne par les
``machinistes''. Au d\'ebut j'ai \'et\'e \'etonn\'e de d\'ecouvrir
qu'\`a Saturne le ``Groupe Th\'eorie'' \'etait constitu\'e uniquement de gens
qui faisaient de la recherche sur le fonctionnement de la machine tandis qu'il 
n'y avait pour ainsi dire pas de th\'eoriciens de physique
nucl\'eaire et des particules
au laboratoire. Ceci indiquait les priorit\'es des fondateurs du laboratoire
et toutes les autres machines que je connais bien doivent se satisfaire de
beaucoup moins d'assistance dans ce domaine. Tout en admettant que les succ\`es
\`a Indiana ont \'et\'e remarquables, on a constat\'e que les particules
polaris\'ees sont difficiles \`a acc\'el\'erer aussi bien \`a Uppsala
qu'\`a J\"ulich. Dans son discours, Lagniel a bien mis en lumi\`ere la
collaboration entre les machinistes et les exp\'erimentateurs, surtout ceux du
groupe nucl\'eon-nucl\'eon, et il a montr\'e comment les deux parties en ont
b\'en\'efici\'e. C'est \`a dire, ce n'\'etait pas toujours la guerre!

Le groupe nucl\'eon-nucl\'eon \'etait le plus gourmand pour les protons
polaris\'es et bien souvent dans le Comit\'e des Exp\'eriences on
savait qu'il y avait toujours un coll\`egue tch\`eque qui \'etait
pr\^et \`a prendre le faisceau
pendant n'importe quelle p\'eriode de vacances. C'\'etait peut-\^etre une
co\"{\i}ncidence, mais au moment o\`u ils ont termin\'e leur programme sur la
diffusion \'elastique proton-proton et proton-neutron, les autorit\'es
fermaient le laboratoire! Les \'etudes nucl\'eon-nucl\'eon \'etaient
peut-\^etre les recherches les plus fondamentales de tout de ce
qu'on a fait \`a Saturne. Cela a \'et\'e fait d'une mani\`ere tr\`es
professionnelle et dans plusieurs cas le groupe a suffisamment
de donn\'ees pour faire
des reconstructions directes des amplitudes, sans passer par des
mod\`eles th\'eoriques.

Saturne \'etait vraiment unique dans le domaine des faisceaux de deutons
combinant une haute intensit\'e avec de grandes valeurs de polarisations
vectorielles \underline{et} tensorielles. Sans Saturne nous n'aurions pas
pu calibrer les polarim\`etres tensoriels de deutons, AHEAD~\cite{AHEAD},
 et POLDER~\cite{POLDER}. Comme cela a \'et\'e  montr\'e par Gar\c{c}on dans 
son expos\'e, ces polarim\`etres ont jou\'e un r\^ole crucial dans la
s\'eparation des facteurs de forme du deuton par la mesure de la polarisation
du deuton de recul dans la diffusion \'elastique \'electron-deuton. Le Comit\'e
des Exp\'eriences et la Direction de Saturne ont toujours eu une vision tr\`es
large de sorte que
le progr\`es de la physique \'etait le but essentiel du laboratoire, et m\^eme
s'il ne s'agissait que de la calibration d'\'equipements pour faire de
\underline{bonnes} exp\'eriences ailleurs qu'\`a Saturne, alors le prix valait
d'\^etre pay\'e. Cela a s\^urement pay\'e dans le cas du facteur de
forme du deuton. Michel Gar\c{c}on nous a montr\'e les r\'esultats
pr\'eliminaires de CEBAF sur T$_{20}$ en $ed\to e\vec{d}$. La collaboration
a pu travailler \`a des valeurs de $q^2$ deux fois plus \'elev\'ees que celles
des exp\'eriences pr\'ec\'edentes avec une efficacit\'e bien plus grande que
celles de toutes les autres m\'ethodes concurrentes.

Le polarim\`etre POLDER se base sur la r\'eaction d'\'echange de charge du
deuton $\vec{d}p \to \{pp\}n$, qui pr\'esente, selon les pr\'edictions, un
signal de pouvoir d'analyse important si la paire finale proton-proton a une 
\'energie d'excitation faible et si elle est produite avec un faible transfert
d'impulsion par rapport au deuton incident~\cite{BW}. Cette r\'eaction a
\'et\'e \'etudi\'ee dans le domaine de quelques centaines de MeV~\cite{EMRIC},
mais \'egalement dans le cas des deutons de 1--2~GeV avec la remarquable
technique qui consiste \`a d\'etecter les deux protons dans le plan focal
du spectrom\`etre SPESIV~\cite{Gaarde1}.  Cette \'etude faisait partie d'une
longue s\'erie d'exp\'eriences d'\'echange de charge
avec diff\'erentes sondes effectu\'ees pendant des ann\'ees par le groupe de
Copenhague-Orsay, et c'est une grande tristesse pour nous tous que le chef du
groupe, Carl Gaarde, nous a quitt\'es quelques semaines avant cette r\'eunion.

Mich\`ele Roy-St\'ephan nous a rappel\'e la quantit\'e \'enorme de
r\'esultats obtenus en 14 ans, surtout en ($^3$He,t), et a insist\'e sur
l'importance de la collaboration active avec les th\'eoriciens pour
l'intepr\'etation de ces r\'esultats. Une collaboratrice parmi d'autres,
Madeleine Soyeur, nous a montr\'e comment de tels r\'esultats de physique
pouvaient \^etre excitants pour les th\'eoriciens. Je crois bien qu'elle a
communiqu\'e cet enthousiasme \`a tout le monde  dans son discours.

La m\'ethode de d\'etection des deux protons dans SPESIV a m\^eme pu \^etre
adapt\'ee \`a l'\'etude de la diffusion profond\'ement in\'elastique dans
laquelle l'isobare $\Delta$ est produit sur l'hydrog\`ene ou sur des
cibles nucl\'eaires~\cite{Gaarde2}. La r\'eaction $(\vec{d},2p)$ est une sonde
int\'eressante pour les transitions $\Delta T = \Delta S =1$, mais inspir\'e
par Marcel Morlet, le groupe
d'Orsay a montr\'e que les mesures du transfert de spin vectoriel en
$(\vec{d},\,\vec{d}^{\,'})$ peuvent fournir des signaux utiles sur les \'etats
$\Delta T=0$, $\Delta S =1$ dans le noyau r\'esiduel~\cite{Morlet}. Il
fallut pas mal d'intuition pour trouver cela
car la sym\'etrie utilis\'ee ici est seulement une
approximation et une s\'eparation compl\`ete aurait exig\'e la combinaison d'un
faisceau polaris\'e tensoriellement et d'un polarim\`etre tensoriel.

De la m\^eme importance permanente \'etait la grande s\'erie de mesures
destin\'ee \`a explorer le probl\`eme \`a petit nombre de nucl\'eons dans
le domaine des \'energies interm\'ediaires. Les mesures \`a triple spin dans
$\vec{d}\vec{p}\to\vec{p}d$ \`a 1.6~GeV~\cite{Igo} repr\'esentaient un vrai
tour de force qui ne sera probablement jamais r\'ep\'et\'e. Plus dr\^oles par
contre sont les exp\'eriences exploratoires qui mesurent une seule observable
en fonction de l'\'energie, par exemple le pouvoir d'analyse tensoriel T$_{20}$
pour le deuton dans $\vec{d}p \to pd$ vers l'arri\`ere~\cite{Arvieux} ou bien
dans la production de pions par la r\'eaction $\vec{d}p\to\,^{3}$He$\,\pi^0
$ \cite{Kerboul,Nikulin}.

A cause de la composante D, le deuton a une d\'eformation g\'eom\'etrique qui
lui donne une forme de cigare dans l'espace des coordonn\'ees, mais qui
ressemble
plut\^ot \`a une cr\^epe dans l'espace des impulsions. Ceci a pu \^etre
etudi\'e en mesurant la d\'ependance en T$_{20}$ du
 mouvement de Fermi dans la cassure du deuton $\vec{d}p
\to ppn$~\cite{Punjabi1}, mais la richesse de Saturne en polarim\`etres  
permit aussi la mesure de la polarisation du proton final~\cite{Stan}.

L'\'etude des \'etats D dans le $^6$Li~\cite{Punjabi2} a \'et\'e r\'ealis\'ee
dans des mesures analogues de cassure, avec des faisceaux d'ions polaris\'es de
plusieurs GeV/c (le moment doit \^etre \'elev\'e par rapport au mouvement de
 Fermi) et ceci semble \^etre la seule exp\'erience \underline{publi\'ee}
utilisant un
tel faisceau. Le synchrotron MIMAS fut con\c{c}u pour donner des faisceaux
polaris\'es de haute qualit\'e et \'egalement des faisceaux d'ions jusqu'au
$^{129}$Xe (30$^{+}$), mais la demande pour des ions aussi lourds n'a
repr\'esent\'e que 10\% du total des demandes pendant des ann\'ees. En partie
cela pourrait \^etre d\^u \`a une grande concurrence avec d'autres facilit\'es
fournissant des ions lourds, comme Ganil et SIS, tandis que dans le domaine des
ions l\'egers, Saturne r\'egnait de fa\c{c}on supr\^eme. Sauf pour le tonneau
de Diog\`ene, Saturne manquait de d\'etecteurs sp\'ecifiques pour les ions
lourds et les
groupes avaient tendance \`a venir s'installer avec leurs propres petits
\'equipements pour l'\'etude de la multi-fragmentation ou pour d'autres
\'etudes~\cite{Gelbke}. Saturne a eu le grand
succ\`es que l'on conna\^{\i}t dans les ions l\'egers gr\^ace \`a toute sa
collection de
spectrom\`etres remarquables qui \'etaient faits sur mesure pour ce travail. 

J'ai d\'ej\`a dit que la physique doit \^etre excitante et aucune revue sur
le travail \`a Saturne ne serait compl\`ete sans la mention du
``Pari de Pascal'',
 et l'ambition humaine de trouver quelque chose de \underline{vraiment}
passionnant, m\^eme si la probabilit\'e d'y arriver est infiniment petite.
S'il y avait un bon joueur de bridge dans la salle, il pourrait
vous dire qu'il faut toujours penser \`a la d\'efense; dans un tournoi
international il ne devrait jamais \^etre question de laisser \`a l'adversaire 
la chance de faire un trop bon ``score''. Dans ce sens, Saturne ne pouvait
permettre que des d\'ecouvertes qui boulverseraient le monde dans son domaine
de comp\'etence soient faites ailleurs, s'il y avait la possibilit\'e de les
faire \`a Saclay. Dans d'autres domaines cette maladie se manifesta par des
tentatives visant \`a r\'ep\'eter les d\'ecouvertes de Pons et Fleischmann sur 
la fusion froide, mais \`a Saturne c'\'etait la chasse aux r\'esonances
dibaryoniques \'etroites\cite{Boris1}, aux \'etats li\'es pion-nucl\'eon
~\cite{Boris2}, \`a la production anormale de pions~\cite{Julien}, aux
anomalons~\cite{anomalon} {\it etc.}. C'\'etait vraiment comme dans ``Les
vacances de M. Hulot''. Rien de tr\`es convaincant n'a jamais \'et\'e trouv\'e
qui ne fut \`a la limite des barres d'erreurs syst\'ematiques ou statistiques.
Mes coll\`egues seront soulag\'es d'apprendre que cette approche est maintenant
poursuivie dans les laboratoires qui succ\`edent \`a Saturne. Je n'ai pas
non plus r\'eussi dans les PACs de COSY et CELSIUS \`a faire arr\^eter la
recherche du $d'$ et ce manque de succ\`es de ma part est la suite
pr\'evisible de mes \'echecs avec les dibaryons \'etroits aupr\`es du Comit\'e
des Exp\'eriences \`a Saturne.

Il est vrai qu'une telle d\'ecouverte dans ce domaine aurait pu cr\'eer une
certaine d\'efense de Saturne et ce sont souvent ces recherches qui
\'eveillent
ce qu'il y a de meilleur dans l'ing\'eniosit\'e technique. Une illustration en
est la fameuse roue de Beurtey et Saudinos~\cite{Beurtey}. Pour effectuer des
mesures simultan\'ees de $A_y$ en diffusion proton-proton \'elastique pour
un grand nombre d'\'energies diff\'erentes, ils ont construit un d\'egradeur
d'\'energie avec beaucoup de marches, qui a un peu l'allure des escaliers
circulaires d'Escher. En faisant tourner rapidement cette roue dans le
faisceau, il \'etait possible d'obtenir des donn\'ees de
haute statistique simultan\'ement \`a 16 \'energies diff\'erentes.
Quelques postes furent suffisants pour tuer un dibaryon, mais ce n'\'etait 
pas un dibaryon fran\c{c}ais!
 
Pendant des ann\'ees la valeur de Saturne pour la calibration de nombreux
autres \'equipements en plus des polarim\`etres a \'et\'e inestimable. 
Paradoxalement on ne se souviendra pas beaucoup de Saturne pour ce
type de travail dans les publications de physique qui seront effectu\'es 
d'ici quelque temps dans d'autres
laboratoires. En revanche les gens vont se souvenir de Saturne quand ils
r\'ealiseront qu'ils ne peuvent plus l'utiliser pour tester des compteurs
pour LHC par exemple! Souvent la recherche appliqu\'ee \`a Saturne donnait
de nouvelles et int\'eressantes possibilit\'es dans des domaines de physique
voisins. J'ai en t\^ete le travail \'elabor\'e par Rolf Michel sur les
pseudo-m\'et\'eorites.

Vous aurez remarqu\'e que je n'ai pas parl\'e des trois grands programmes qui
ont domin\'e les deux derni\`eres ann\'ees du fonctionnement de Saturne-2,
{\it viz} DISTO, SPESIV$\pi$, et la Transmutation Nucl\'eaire. Il est bien
trop t\^ot pour en juger la signification. Nous esp\'erons tous que
les informations fournies par les exp\'eriences sur la transmutation se
montreront pr\'ecieuses  dans la qu\^ete d'une m\'ethode fiable pour traiter
les d\'echets nucl\'eaires. Ceci sera un vrai monument pour les travaux de
Saturne, peut-\^etre un monument plus important que celui de Ptol\'em\'ee
qu'on vient de d\'evoiler devant le Grand Palais.\\

Pour terminer, je dois remercier Alain, Fran\c{c}oise, Simone et Bernard
qui ont assur\'e le succ\`es de cette r\'eunion.
Il est bien \'evident que le fran\c{c}ais n'est la langue maternelle ni de
M.~Blair ni de moi. N\'eanmoins il va rester ma langue fraternelle et cela est
pour une grande part d\^u aux collaborations fructueuses avec des gens de
Saturne, beaucoup trop nombreux pour \^etre cit\'es.\\

Saturne, nous allons tous nous souvenir de toi.  Saturne, adieu! Merci!

\end{document}